\documentclass[12pt]{article}
\usepackage{pic02}
\usepackage{hyperref}
\usepackage{graphicx}

\begin{document}

\title{\bf RESULTS ON $\sin 2\phi_2 (\alpha)$ FROM THE B FACTORIES} 
\author{Thomas E. Browder         \\
{\em University of Hawaii at Manoa}}
\maketitle

%
% photograph of author
%  This is where we will insert a photograph. To see what it would look like,
%  uncomment the following lines.
%
%\begin{figure}[h]
%\begin{center}
%
% include photograph for proceeding version
%
%\includegraphics[height=4.5cm]{einstein.eps}
%
% insert a fixed vertical spacing instead for the ArXiv preprint
%
\vspace{4.5cm}
%
%\end{center}
%\end{figure}

\baselineskip=14.5pt
\begin{abstract}
The status of CP violation in $B^0\to \pi^+\pi^-$ and the
determination of $\sin 2\phi_2 (\alpha)$ from the $B$ factories
is described. 
\end{abstract}
\newpage

\baselineskip=17pt

\section{Introduction}

In 1973 Kobayashi and Maskawa (KM) proposed a model
where $CP$ violation is
incorporated as an irreducible complex phase in the
weak-interaction quark mixing matrix~\cite{KM}.
Recent measurements of the $CP$-violating parameter $\sin 2\phi_1$ 
by the Belle~\cite{Belle_CPV2} and BaBar~\cite{BaBar_CPV2} collaborations 
have clearly established $CP$ violation in the neutral $B$ meson system that
is consistent with KM expectations. The next step in the program
is measurements of other $CP$-violating parameters.
Here we describe recent measurements of $CP$-violating
asymmetries in the mode $B^0 \to \pi^+\pi^-$~; 
these are sensitive to the parameter 
$\sin 2\phi_2$ (also known as $\sin 2\alpha$).

\section{Experimental Challenges}

To measure $\phi_2$ (a.k.a $\alpha$), the two most promising
approaches involve the use of the decay modes $B^0\to \pi^-\pi^+$
and $B^0\to \rho^{\pm}\pi^{\mp}$. The former is an example of 
a CP eigenstate and is thus the most straightforward approach
as well as the mode with the best senstivity.
The interference in the $B\to \pi^+\pi^-$ mode
between the direct decay and the decay
via mixing leads to a 
CP violating asymmetry with a sin-like time modulation
as in charmonium CP eigenstate modes such as $B^0\to \psi K_s$.

The KM model predicts sizeable
$CP$-violating asymmetries in the time-dependent
rates for  $B^0$ and $\bar{B}^0$
decays to a common $CP$ eigenstate, $f_{CP}$.
In the decay chain $\Upsilon(4S)\to B^0 \bar{B}^0 \to f_{CP}f_{\rm tag}$,
where one of $B$ mesons decays at time $t_{CP}$ to $f_{CP}$ 
and the other decays at time $t_{\rm tag}$ to a final state
$f_{\rm tag}$ that distinguishes between $B^0$ and $\bar{B}^0$, 
the decay rate has a time dependence given by
\begin{eqnarray}
\label{eq:R_q}
{\cal P}_{\pi\pi}^q(\Delta{t}) = 
\frac{e^{-|\Delta{t}|/{\tau_b}}}{4{\tau_b}}
\left[1 + q\cdot 
\left\{ {\cal S}_{\pi\pi}\sin(\Delta {m_d}\Delta{t})   \right. \right. \nonumber \\
\left. \left.
   + {\cal A}_{\pi\pi}\cos(\Delta {m_d}\Delta{t})
\right\}
\right],
\end{eqnarray}
where $\tau_b$ is the $B^0$ lifetime, $\Delta {m_d}$ is the mass difference 
between the two $B^0$ mass
eigenstates, $\Delta{t}$ = $t_{CP}$ $-$ $t_{\rm tag}$, and
the $b$-flavor charge $q$ = +1 ($-1$) when the tagging $B$ meson
is a $B^0$ ($\bar{B}^0$).
The $CP$-violating parameters, $S_{\pi\pi}$ and $A_{\pi\pi}$, 
defined in Eq.~(\ref{eq:R_q}) can be expressed by
in terms of the complex parameter
$\lambda$ that depends on both $B^0-\bar{B}^0$
mixing and on the amplitudes for $B^0$ and $\bar{B}^0$ decay to 
$\pi^+\pi^-$\cite{lambda}. In the Standard Model, to a good approximation,
$|\lambda|$ is equal to the absolute value
of the ratio of the $\bar{B}^0$ to $B^0$ decay amplitudes.

The experimental technique is similar to that used for the
$\sin 2\phi_1 (\beta)$ measurement \cite{pan_cpv}.
However, there are several additional complications and differences.
The decay amplitude for $B^0\to \pi^+\pi^-$ contains a contribution
from a tree diagram ($b\to u \bar{u} d$) as well as a Cabibbo
suppressed penguin diagram ($b\to s \bar{u} s$). The penguin
contribution is not negligible and has a weak phase that is
different from the phase of the larger tree amplitude, which is
zero in the usual parameterization. In general, the penguin
contribution will also have a strong phase.
Therefore the time dependent asymmetry, proportional to
$\sin(\Delta m \Delta t)$ and parameterized by $S_{\pi\pi}$, 
which is measured is not equal to $\sin 2\phi_2$ but
instead will have a large unknown correction.
The presence of the extra contribution also induces an additional
time dependent term proportional to $\cos(\Delta m \Delta t)$,
parameterized by $A_{\pi\pi}$\cite{convention}. 
This is called {\it penguin pollution}.
As the notation $A_{\pi\pi}$ suggests, the asymmetry term with
$\cos(\Delta m \Delta t)$ modulation is due to direct
CP violation. Note that unlike the mixing induced CP violation,
the direct CP violation term does not time integrate to zero.

There are a number of other purely experimental complications. The
branching fraction for the $B^0\to \pi^+\pi^-$ decay is quite
small (see Table~\ref{table_bf})
compared to the charmonium modes, only $(4.8\pm 0.6)\times
10^{-6}$. Thus, very large data samples are required. The BaBar
results are based on a sample of $88\times 10^6$ $B\bar{B}$ pairs.
Belle has recorded a sample of comparable size, but has published
results with a subset of $45\times 10^{6}$ $B\bar{B}$ pairs.

The other challenging
requirement for the detector is the separation
of kaons from pions at high momentum. This is needed to distinguish
$\bar{B}^0\to \pi^+ \pi^-$ from $\bar{B}^0\to K^-\pi^+$,
which has similar kinematics and
a branching fraction about three times larger. 
Two approaches to high momentum particle identification have been
implemented at the $B$ factory experiments.
 Both are based on the use of Cerenkov radiation.

At Belle, aerogel Cerenkov radiators are used. Blocks of
aerogel are readout by fine-mesh phototubes that have
high-gain and operate comfortably in a 1.5 Tesla magnetic field. 
Since the threshold for the aerogel is around 1.5 GeV, below this
momentum K/$\pi$ separation is carried out using high precision
time-of-flight
scintillators with resolution of 95 ps. The aerogel and TOF counter
system are complemented by dE/dx measurements in the central drift
chamber. The dE/dx system provides additional 
K/$\pi$ separation around 2.5 GeV
in the relativistic rise region as well as below 0.7 GeV.
For high momentum kaons, an efficiency of 88\% with a
misidentification probability below 9\% has been achieved.

At BaBar, Cerenkov light is produced in quartz bars and then
transmitted
by total internal reflection outside the detector through a water tank
to a large array of phototubes
where the ring is imaged. The detector is referred to by the acronym
DIRC. It provides 
$K/\pi$ separation that ranges from $8 \sigma$ at 2 GeV to
$2.5\sigma$ at 4 GeV. 

\begin{table}
\centering
\caption{ \it Branching Fractions in units of $10^{-6}$
for $B\to K \pi$ and $B\to \pi\pi$ Modes.}
\vskip 0.1 in
\begin{tabular}{|l|c|c|c|} \hline
          &  BaBar & Belle & CLEO \\
\hline
\hline
 $B^0\to \pi^+\pi^-$  & $ 4.6 \pm 0.6\pm 0.2$    &
             $ 5.4\pm 1.2 \pm 0.5$ & $ 4.3^{+1.6}_{-1.4}\pm 0.5$     \\
 $B^+\to \pi^+\pi^0$  & $ 5.5^{+1.0}_{-0.9}\pm 0.6$    &
             $ 7.4\pm 2.2 \pm 0.9$   & $5.4\pm 2.6 $          \\
 $B^0\to K^{\pm}\pi^{\mp}$  & $17.9\pm 0.9\pm 0.7 $  &
             $22.5\pm 1.9\pm 1.8 $  & $ 17.2^{+2.5}_{-2.4}\pm 1.2 $ \\
 $B^+\to K^+\pi^0$  & $ 12.8^{+1.2}_{-1.1}\pm 1.0$    &
 $ 13.0^{+2.5}_{-2.4} \pm 1.3$  & $11.6^{+3.0+1.4}_{-2.7-1.3} $    \\
 $B^+\to K^0\pi^+$  & $ 17.5^{+1.8}_{-1.7}\pm 1.3 $    &
$ 19.4^{+3.1}_{-3.0} \pm 1.6 $   & $18.2^{+4.6}_{-4.0}\pm 1.6$ \\
 $B^0\to K^0\pi^0$  & $ 10.4\pm 1.5\pm 0.8 $    &
$ 8.0^{+3.3}_{-3.1} \pm 1.6$   & $14.6^{+5.9+2.4}_{-5.1-3.3}$\\
\hline
\end{tabular}
\label{table_bf}
\end{table}

Hoewever, even after 
the application of high momentum particle identification,
the $B^0\to \pi^+\pi^-$ 
CP eigenstate signal sits on a very large continuum background
from  $e^+e^- \rightarrow q\overline{q}$ ($q = u,~d,~s,~c$) processes.
Several analysis techniques to reduce this background have been
developed.

BaBar uses a selection on the angle between the sphericity axis
of the $B$ candidate and the sphericity axis of the rest of the event
(denoted $\theta_S$).
The cosine of this angle is uniformly distributed for the B signal
and is concentrated at $\cos\theta_S=\pm 1$ for continuum
background. After requiring that $|\cos(\theta_S)|<0.8$, they form
a Fisher discriminant, $F$, from the energies in nine cones of increasing
angular aperture opposite the $B$ candidate. No cut is applied,
instead $F$ is used a 
variable to distinguish signal from continuum in their fit.
This technique was originally developed by CLEO.

Belle uses a likelihood based technique
in order to suppress continuum background. 
Signal and background
likelihood functions, ${\cal L}_S$ and ${\cal L}_{BG}$, are formed
from two variables. One is a Fisher
discriminant determined from six modified Fox-Wolfram
moments~\cite{SFW}; the other is 
the $B$ flight direction in the cms, 
with respect to the $z$ axis ($\cos\theta_B$).
The signal likelihood ${\cal L}_S$ is determined from Monte Carlo (MC)
and ${\cal L}_{BG}$ from data, and
${\cal L}_S/({\cal L}_S+{\cal L}_{BG}) > 0.825$ is required 
for candidate events.

\begin{figure}[htb]
\begin{center}
\includegraphics[width=8cm]{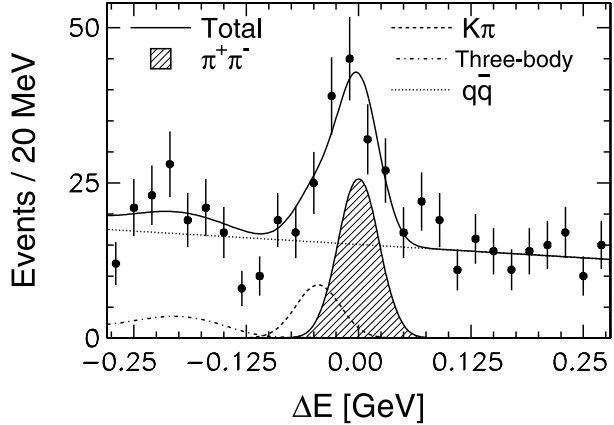}
\end{center}
 \caption{\it
      Belle data:  $\Delta E$ distribution for $\pi^+\pi^-$ 
event candidates that are in the
 $M_{\rm bc}$ signal region.
    \label{belle_fig1} }
\end{figure}
\begin{figure}[htb]
\begin{center}
\includegraphics[width=7cm]{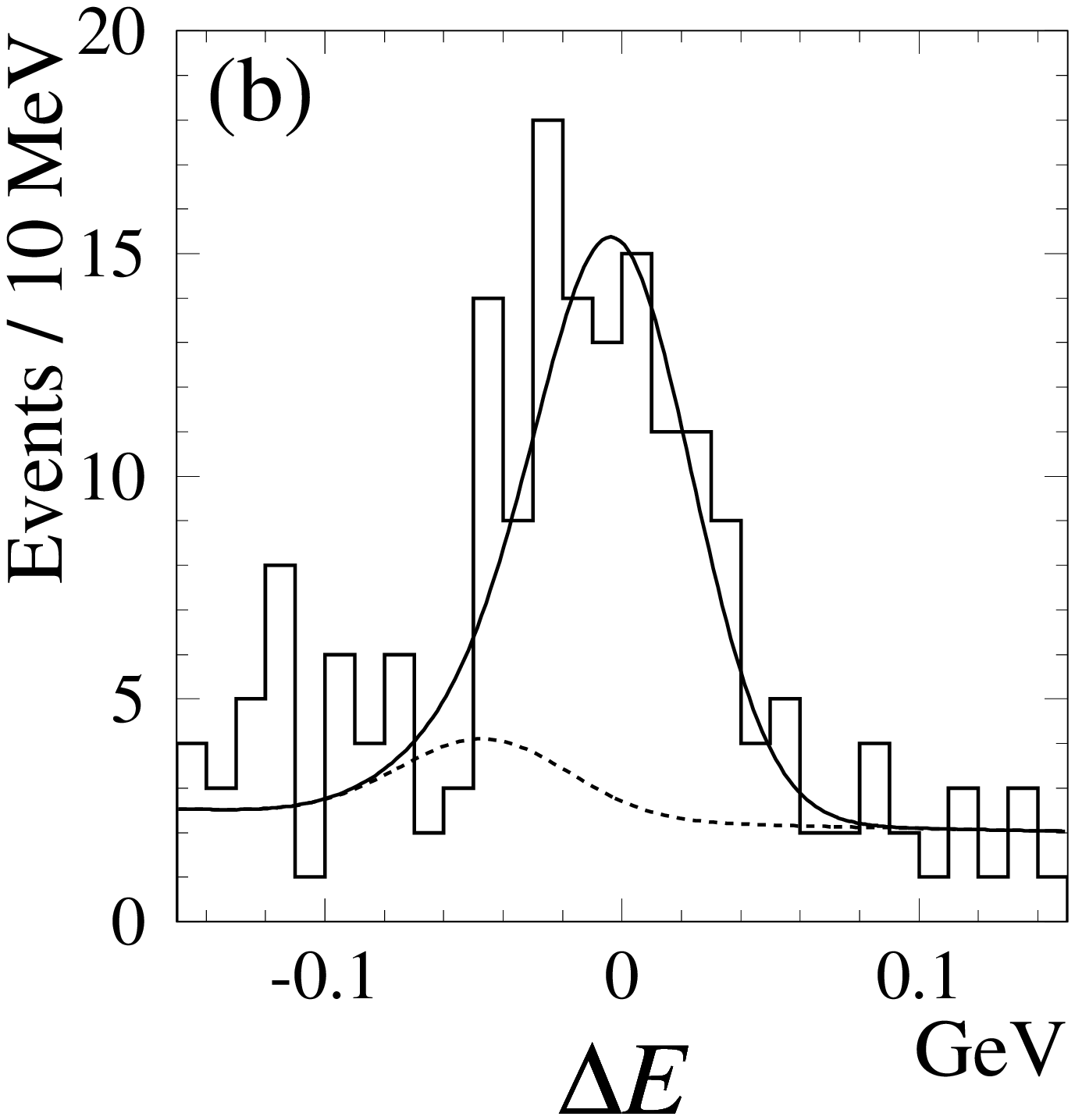}
\end{center}
 \caption{\it
      BaBar data: The $\Delta E$ distribution for events
     enhanced in $\pi\pi$ signal by cuts. 
     The solid curves represent the projection
     of the maximum likelihood fit while the dashed curves represent
     the sum of $q \bar{q}$ background 
     and misidentified $K^{\pm}\pi^{\mp}$ events.
    \label{babar_fig1} }
\end{figure}

The signal to continuum background ratio 
is a strong function of the tagging method; this
effect must be taken into account in the CP extraction. 
There is also still some residual
background from misidentified $B^0\to K^+\pi^-$ as well,
although this background is reasonably well separated by the kinematic
variable $\Delta E$.

After flavor tagging and vertexing requirements are applied,
a likelihood fit is applied to extract the two CP violation parameters.
At Belle, an unbinned fit to the $\Delta t$ distribution of 162
candidates in the signal region is applied. The signal of 
$73.5\pm 13.8$ events is shown in Fig.~\ref{belle_fig1}.
The signal to background
fraction is a function of tagging purity and divided
into six bins. The only free
parameters in the Belle fit are $S_{\pi\pi}$ and $C_{\pi\pi}$.
At BaBar, a more complex fit to $m_{ES}$, $\Delta E$, $F$ (the event
shape Fisher discriminant), Cerenkov angles $\theta_c^+$, $\theta_c^-$,
 and $\Delta t$ is performed for a sample of 26070 events
of which $157\pm 19\pm 17$ are signal events. A signal enhanced
$\Delta E$ distribution is  shown for a
sub-sample in Fig.~\ref{babar_fig1}.
 The fit has a total of 76
parameters. These include the values of $S_{\pi\pi}$ and
$A_{\pi\pi}$ (2); signal and background yields (5); $K \pi$ charge
asymmetries (2); signal and background tagging efficiencies (16)
and efficiency asymmetries (16); signal mistag fraction and mistag
fraction differences (8); signal resolution function (9); 
and parameterization of background shapes in $m_{ES}$(5), $\Delta E$(2),
$F$(5) and $\Delta t$(6). This somewhat more complex approach
has good statistical reach.  However, the background must be
accurately parameterized since events with rather poor signal
to background ratios (O(1/10)) are used.

To validate the analysis, a variety of consistency checks are
performed. For example, both BaBar and Belle measure 
the $B$ lifetime and mixing
frequency in the $B^0\to K^-\pi^+$ sample. They find results
consistent with the world averages. A variety of control samples
are also examined. For instance, Belle takes $D^{(*)+}\pi^-$ events,
adds additional background from the $B\to \pi\pi$ sidebands
to degrade the signal to background ratio to the level of the
$\pi\pi$ signal, and then performs the CP fit. They find 
$A_{\pi\pi}=0.03\pm 0.04$ and $S_{\pi\pi}=0.08\pm 0.06$.
No artificial CP asymmetries are found in any of the control samples
that have been studied.

\section{Results}
\begin{figure}[htb]
\begin{center}
\includegraphics[width=7cm]{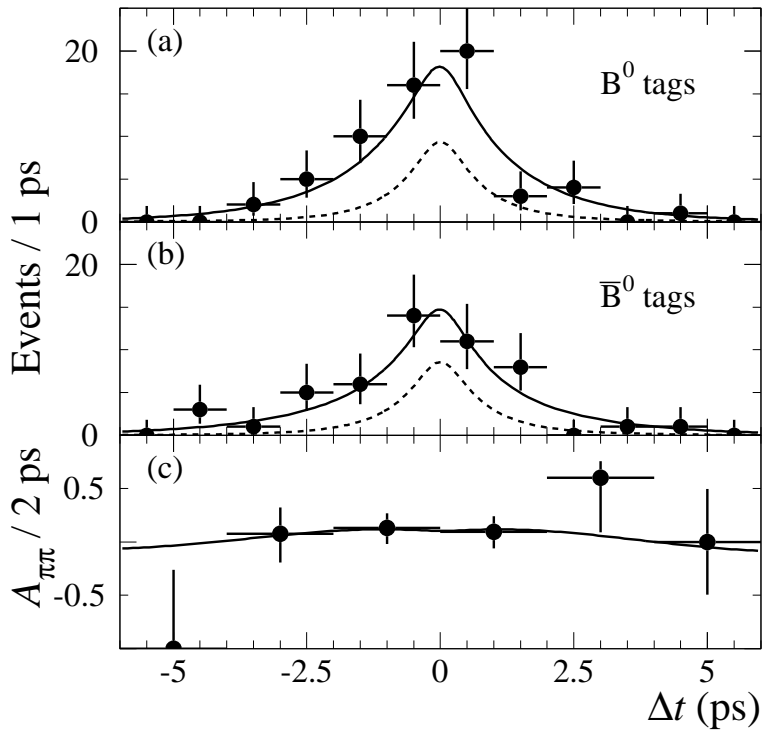}
\end{center}
 \caption{\it
      BaBar data: Distributions of $\Delta t$ for events in signal
      enhanced in signal $\pi\pi$ decays with (a) $B^0$ tags 
     (b) $\bar{B}^0$ tags and 
     (c) the asymmetry $A_{\pi\pi}(\Delta t)$ as a function of
     $\Delta t$. The solid curves represent the projections of
     the maximum likelihood fit, dashed curves represent the sum
     of $q \bar{q}$ and background events.
    \label{babar_fig2} }
\end{figure}

The observed flavor tagged $\Delta t$ and asymmetry 
distributions in BaBar data with cuts to enhance the signal fraction
are shown in Fig.\ref{babar_fig2}.
No sin-like modulation is observed in the asymmetry distribution
while there is a slight hint of a cos-like term. 
For the CP parameters, $A_{\pi \pi}$, 
BaBar obtains 
\begin{eqnarray}
A_{\pi\pi}= 0.30\pm 0.25 \pm 0.04 \\
S_{\pi\pi}= 0.02\pm 0.34 \pm 0.05
\end{eqnarray}
From these results,
BaBar obtains  90\% confidence level intervals
for $A_{\pi\pi}$ of $[-0.12,0.72]$ and for $S_{\pi\pi}$
of $[-0.54, 0.58]$.

\begin{figure}[htb]
\begin{center}
\includegraphics[width=9.5cm]{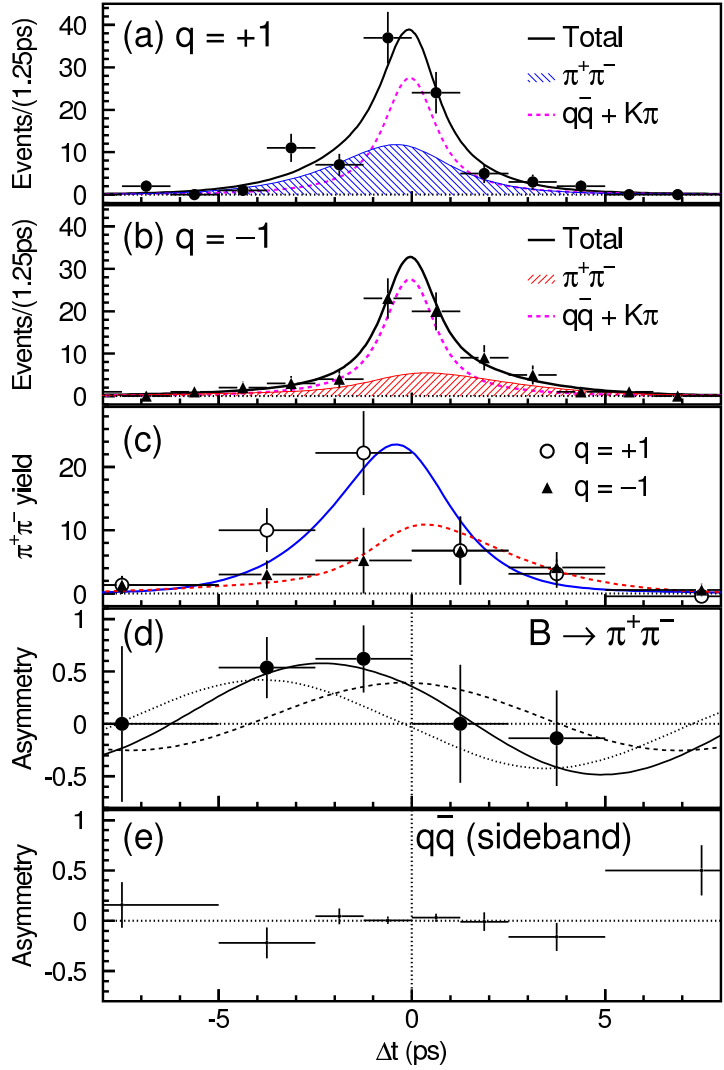}
\end{center}
 \caption{\it
      Belle data: the $\Delta t$ and asymmetry distributions 
 for the $B^0 \rightarrow \pi^+\pi^-$ candidates: 
(a) candidates with $q = +1$, i.e. the tag side is identified as
 $B^0$;
(b) candidates with $q = -1$;
(c) $\pi^+ \pi^-$ yields after background subtraction. 
(d) the $CP$ asymmetry for $B^0 \rightarrow \pi^+\pi^-$
after background subtraction. The
point in the rightmost bin has a large negative value
that is outside of the range of the histogram;
(e) the raw asymmetry for $B^0 \to \pi^+\pi^-$ sideband events.
In Figs. (a) through (c), the curves show the
results of the unbinned maximum likelihood fit.
In Fig. (d), the solid curve shows the resultant $CP$ asymmetry,
while the dashed (dotted) curve is the contribution from
the cosine (sine) term.
    \label{belle_fig2} }
\end{figure}

The Belle $\Delta t$ distributions before and after
background subtractions are shown in Fig.~\ref{belle_fig2}.
The difference in the height of the $B^0$ and $\bar{B^0}$ tags in 
Fig.~\ref{belle_fig2}(c)
is an indication of direct CP violation. The blue and red curves
for $B^0$ and $\bar{B}^0$ tags are also asymmetric in time.
The asymmetry distribution
in Fig.~\ref{belle_fig2}(d)
suggests the presence of sin-like as well
as cos-like modulations.
In contrast to BaBar, Belle finds 
\begin{eqnarray}
~A_{\pi\pi}= 0.94^{+0.25}_{-0.31}\pm 0.09 \\
S_{\pi\pi}= -1.21^{+0.38+0.16}_{-0.27-0.13}
\end{eqnarray}
Each of these two measurements is only $2.9\sigma$ from zero, which
is not yet statistically overwhelming.

\begin{figure}[htb]
\begin{center}
\includegraphics[width=11cm]{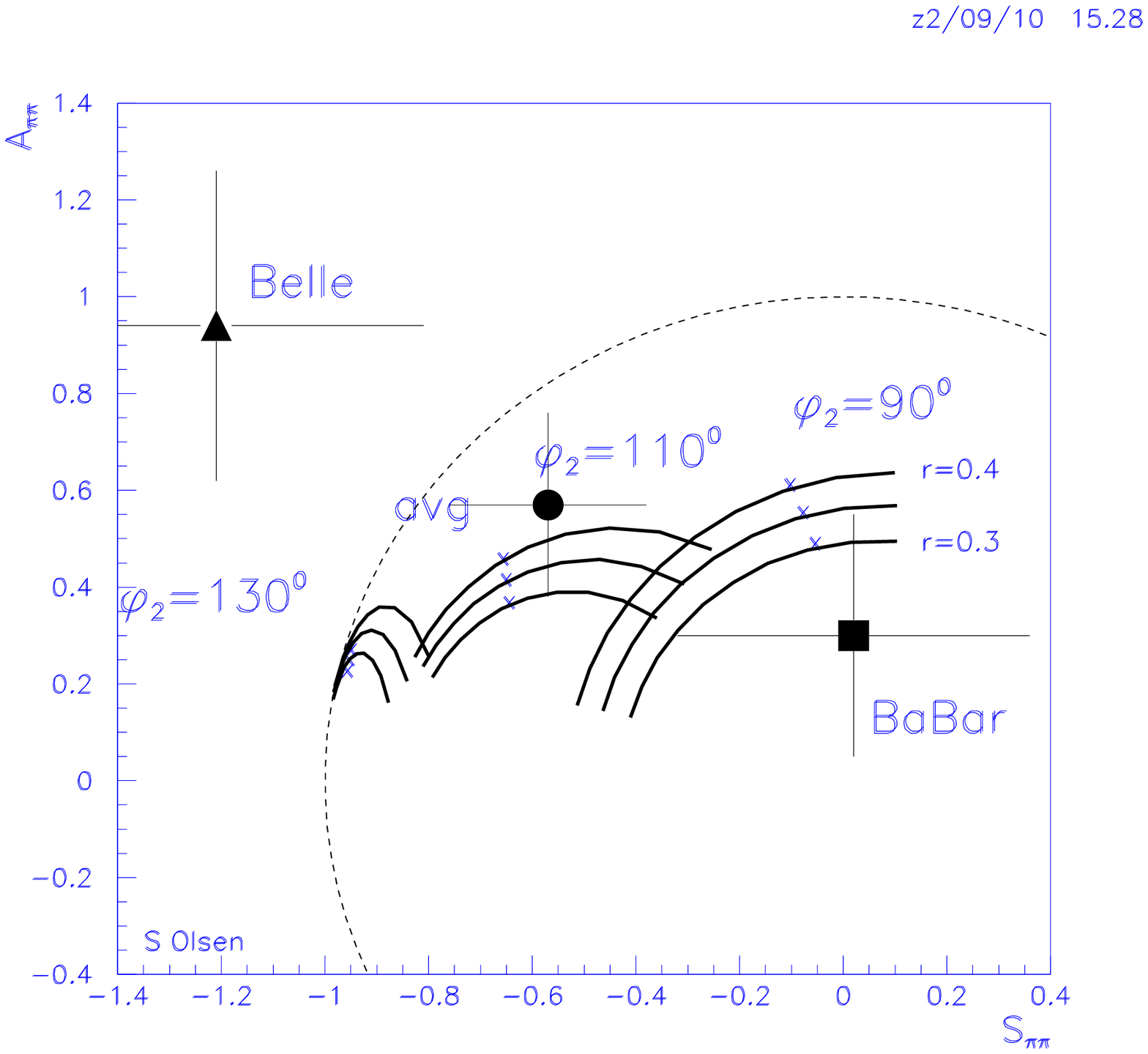}
\end{center}
 \caption{\it
      Comparison of Belle vs BaBar CP Measurements ($S_{\pi\pi}$
      versus $A_{\pi\pi}$) with theoretical
      expectations for different values of $\sin 2\phi_2$, $r=|T/P|$,
      and final state interaction (FSI) phase shifts.
    \label{olsenfig} }
\end{figure}

The two sets of
results give somewhat different pictures of the physics.
The $S_{\pi\pi}$ results are statistically marginally consistent.
Nevertheless we can try to assess the physics content of
the results.
The two measurements and their weighted average 
are shown in the space of $A_{\pi\pi}$ 
and $S_{\pi\pi}$ in Fig.~\ref{olsenfig}. This figure also shows
the physical boundary $A_{\pi\pi}^2+S_{\pi\pi}^2=1$. The Belle
measurement is 1.3$\sigma$ from the physical boundary, consistent
with a statistical fluctuation.
The curves in Fig.~\ref{olsenfig}
correspond to different values of $\phi_2$ and
to $r$, the ratio of tree to penguin amplitudes. A given theoretical
curve corresponds to the range of possible FSI phases. The ratio
r is determined from data on $B^+\to \pi^+\pi^0$, $B^+\to K^0_S\pi^+$ 
and $B\to \pi\ell\nu$\cite{rosner}.

\section{Discussion and related modes}

Gronau and London showed that it is possible to
use isospin invariance and measurements of the 
flavor tagged branching fractions
of all $B\to \pi\pi$ branching fractions to
disentangle the effects of penguin pollution in time 
dependent measurements of asymmetries in $B^0\to \pi^+\pi^-$ and
determine $\sin 2\phi_2$ \cite{gronau}.

As shown in Table~\ref{table_bf}, all the $\pi\pi$ decay modes
except for $\pi^0\pi^0$ have now been measured by Belle and Babar
and are in good agreement with CLEO.
Both $B$-factory experiments have also searched for $B\to \pi^0\pi^0$.
With 29 $fb^{-1}$ Belle finds a 2.4 $\sigma$ hint of a signal 
and sets an upper limit at 90\%
confidence level of $6.4 \times 10^{-6}$. Babar uses their
full dataset and also finds a modest excess. They
set an upper limit of ${\cal B}(B\to \pi^0\pi^0)<3.6\times 10^{-6}$
at the 90\% confidence level. It is {\it possible} that the excesses
seen will become signals with much larger data samples and that the 
$B^0\to \pi^0\pi^0$ branching fraction is of order $1\times 10^{-6}$.

It is also possible to determine ratios of partial widths
of the modes $B^+\to \pi^+\pi^0$ and $B^0\to \pi^+\pi^-$
from the measured yields. Belle finds the ratio of widths
\begin{equation}
{\tau^+ \over \tau_0}
{{(B^0\to \pi^+\pi^-)}\over {2(B^+\to \pi^+\pi^0)}}
= 0.40 \pm 0.15 \pm 0.05 << 1 
\label{ratio}
\end{equation}
The factor of two in the denominator accounts for the $\pi^0$ wavefunction.
A comparable ratio of $0.46\pm 0.11$ is obtained from BaBar data.
The deviation of this ratio from unity indicates either some kind
of interference in $B\to \pi^+\pi^-$ or final state rescattering
or the contribution of other diagrams. This is an important clue
to understanding the $B\to \pi\pi$ system.

\section{Conclusion}

Although the samples of $B\to \pi^-\pi^+$ events are still
relatively small
and the continuum backgrounds are large,
the first round of measurements and attempts
to determine $\sin2\phi_2$ using the $\pi^+\pi^-$ mode 
have been reported\cite{babar_pipicpv},\cite{belle_pipicpv}. 
Unlike the case of charmonium modes, the two experiments do not
agree well. With 88 million $B\bar{B}$ pairs, 
BaBar finds no evidence for indirect CP violation and
a direct CP violation parameter consistent with zero. By contrast,
with 45 million $B\bar{B}$ pairs,
Belle finds indications for both indirect and direct CP violation
in the $B\to\pi^+ \pi^-$ system. Belle plans to update their result with
an additional 36 fb$^{-1}$ ($\sim 39\times 10^6$ $B\bar{B}$ pairs)
in the near future.
However, to fully resolve this discrepancy and precisely
determine $\sin 2\phi_2$, much more
data will be needed.

\section{Acknowledgements}
I wish to acknowledge the essential contributions of my
 colleagues on Belle, BaBar, KEK-B and PEP-II to the work described
 here. I also thank the organizers for a well run conference and
 especially Su Dong for his patience.


\begin{thebibliography}{99}
\bibitem{lambda} The observables $S_{\pi\pi}$ and $A_{\pi\pi}$
can be related to $\lambda$:
 $S_{\pi\pi}=2 {\rm Im}(\lambda)/(|\lambda|^2+1)$ and 
$A_{\pi\pi}=(|\lambda|^2-1)/(|\lambda|^2+1)$.
\bibitem{convention} In BaBar papers a slightly different pair
of observables are used: $S_{\pi\pi}$ and $C_{\pi\pi}$, 
where $C_{\pi\pi} = - A_{\pi\pi}$.
\end{thebibliography}
\end{document}